\documentclass[%
 reprint,
 superscriptaddress,
 amsmath,amssymb,
 aps,
 noeprint,
]{revtex4-2}
\usepackage{hyperref}
\hypersetup{
colorlinks=true,
linkcolor=blue,
urlcolor=black,
citecolor=blue,
anchorcolor=blue
}
\usepackage{graphicx}
\usepackage{svg}
\usepackage{float}
\usepackage{physics}
\usepackage{dcolumn}
\usepackage{bm}
\allowdisplaybreaks 

\begin{document}

\preprint{APS/123-QED}

\title{Efficient decoupling of a non-linear qubit mode from its environment}

\author{F.~Pfeiffer}
\email{frederik.pfeiffer@wmi.badw.de}
\affiliation{Technical University of Munich, TUM School of Natural Sciences, Department of Physics, Garching 85748, Germany}
\affiliation{Walther-Mei\ss ner-Institut, Bayerische Akademie der Wissenschaften, Garching 85748, Germany}
\author{M.~Werninghaus}
\affiliation{Technical University of Munich, TUM School of Natural Sciences, Department of Physics, Garching 85748, Germany}
\affiliation{Walther-Mei\ss ner-Institut, Bayerische Akademie der Wissenschaften, Garching 85748, Germany}
\author{C.~Schweizer}
\affiliation{Fakultät für Physik, Ludwig-Maximilians-Universität München, Schellingstraße 4, D-80799 München, Germany}
\author{N.~Bruckmoser}
\affiliation{Technical University of Munich, TUM School of Natural Sciences, Department of Physics, Garching 85748, Germany}
\affiliation{Walther-Mei\ss ner-Institut, Bayerische Akademie der Wissenschaften, Garching 85748, Germany}
\author{L.~Koch}
\affiliation{Technical University of Munich, TUM School of Natural Sciences, Department of Physics, Garching 85748, Germany}
\affiliation{Walther-Mei\ss ner-Institut, Bayerische Akademie der Wissenschaften, Garching 85748, Germany}
\author{N.~J.~Glaser}
\affiliation{Technical University of Munich, TUM School of Natural Sciences, Department of Physics, Garching 85748, Germany}
\affiliation{Walther-Mei\ss ner-Institut, Bayerische Akademie der Wissenschaften, Garching 85748, Germany}
\author{G.~Huber}
\affiliation{Technical University of Munich, TUM School of Natural Sciences, Department of Physics, Garching 85748, Germany}
\affiliation{Walther-Mei\ss ner-Institut, Bayerische Akademie der Wissenschaften, Garching 85748, Germany}
\author{D.~Bunch}
\affiliation{Technical University of Munich, TUM School of Natural Sciences, Department of Physics, Garching 85748, Germany}
\affiliation{Walther-Mei\ss ner-Institut, Bayerische Akademie der Wissenschaften, Garching 85748, Germany}
\author{F.~X.~Haslbeck}
\affiliation{Technical University of Munich, TUM School of Natural Sciences, Department of Physics, Garching 85748, Germany}
\affiliation{Walther-Mei\ss ner-Institut, Bayerische Akademie der Wissenschaften, Garching 85748, Germany}
\author{M.~Knudsen}
\affiliation{Technical University of Munich,TUM School of Computation, Information and Technology, Department of Computer Science, Garching 85748, Germany}
\affiliation{Walther-Mei\ss ner-Institut, Bayerische Akademie der Wissenschaften, Garching 85748, Germany}
\author{G.~Krylov}
\affiliation{Technical University of Munich, TUM School of Natural Sciences, Department of Physics, Garching 85748, Germany}
\affiliation{Walther-Mei\ss ner-Institut, Bayerische Akademie der Wissenschaften, Garching 85748, Germany}
\author{K.~Liegener}
\affiliation{Technical University of Munich, TUM School of Natural Sciences, Department of Physics, Garching 85748, Germany}
\affiliation{Walther-Mei\ss ner-Institut, Bayerische Akademie der Wissenschaften, Garching 85748, Germany}
\author{A.~Marx}
\affiliation{Walther-Mei\ss ner-Institut, Bayerische Akademie der Wissenschaften, Garching 85748, Germany}
\author{L.~Richard}
\affiliation{Technical University of Munich, TUM School of Natural Sciences, Department of Physics, Garching 85748, Germany}
\affiliation{Walther-Mei\ss ner-Institut, Bayerische Akademie der Wissenschaften, Garching 85748, Germany}
\author{J.~Romeiro}
\affiliation{Technical University of Munich, TUM School of Natural Sciences, Department of Physics, Garching 85748, Germany}
\affiliation{Walther-Mei\ss ner-Institut, Bayerische Akademie der Wissenschaften, Garching 85748, Germany}
\author{F.~Roy}
\affiliation{Technical University of Munich, TUM School of Natural Sciences, Department of Physics, Garching 85748, Germany}
\affiliation{Walther-Mei\ss ner-Institut, Bayerische Akademie der Wissenschaften, Garching 85748, Germany}
\author{J.~Schirk}
\affiliation{Technical University of Munich, TUM School of Natural Sciences, Department of Physics, Garching 85748, Germany}
\affiliation{Walther-Mei\ss ner-Institut, Bayerische Akademie der Wissenschaften, Garching 85748, Germany}
\author{C.~Schneider}
\affiliation{Technical University of Munich, TUM School of Natural Sciences, Department of Physics, Garching 85748, Germany}
\affiliation{Walther-Mei\ss ner-Institut, Bayerische Akademie der Wissenschaften, Garching 85748, Germany}
\author{M.~Singh}
\affiliation{Technical University of Munich, TUM School of Natural Sciences, Department of Physics, Garching 85748, Germany}
\affiliation{Walther-Mei\ss ner-Institut, Bayerische Akademie der Wissenschaften, Garching 85748, Germany}
\author{L.~S\"odergren}
\affiliation{Technical University of Munich, TUM School of Natural Sciences, Department of Physics, Garching 85748, Germany}
\affiliation{Walther-Mei\ss ner-Institut, Bayerische Akademie der Wissenschaften, Garching 85748, Germany}
\author{I.~Tsitsilin}
\affiliation{Technical University of Munich, TUM School of Natural Sciences, Department of Physics, Garching 85748, Germany}
\affiliation{Walther-Mei\ss ner-Institut, Bayerische Akademie der Wissenschaften, Garching 85748, Germany}
\author{F.~Wallner}
\affiliation{Technical University of Munich, TUM School of Natural Sciences, Department of Physics, Garching 85748, Germany}
\affiliation{Walther-Mei\ss ner-Institut, Bayerische Akademie der Wissenschaften, Garching 85748, Germany}
\author{C. A. Riofrío}
\affiliation{BMW Group, Munich, Germany}
\author{S.~Filipp}
\email{stefan.filipp@wmi.badw.de}
\affiliation{Technical University of Munich, TUM School of Natural Sciences, Department of Physics, Garching 85748, Germany}
\affiliation{Walther-Mei\ss ner-Institut, Bayerische Akademie der Wissenschaften, Garching 85748, Germany}

\date{\today}

\begin{abstract}
To control and measure the state of a quantum system it must necessarily be coupled to external degrees of freedom. This  inevitably leads to spontaneous emission via the Purcell effect, photon-induced dephasing from measurement back-action, and errors caused by unwanted interactions with nearby quantum systems. To tackle this fundamental challenge, we make use of the design flexibility of superconducting quantum circuits to form a multi-mode element -- an artificial molecule -- with symmetry-protected modes. The proposed circuit consists of three superconducting islands coupled to a central island via Josephson junctions. It exhibits two essential non-linear modes, one of which is flux-insensitive and used as the protected qubit mode. The second mode is flux-tunable and serves via a cross-Kerr type coupling as a mediator to control the dispersive coupling of the qubit mode to the readout resonator. We demonstrate the Purcell protection of the qubit mode by measuring relaxation times that are independent of the mediated dispersive coupling. We show that the coherence of the qubit is not limited by photon-induced dephasing when detuning the mediator mode  from the readout resonator and thereby reducing the dispersive coupling. The resulting highly protected qubit with tunable interactions may serve as a basic building block of a scalable quantum processor architecture, in which qubit decoherence is strongly suppressed.
\end{abstract}

\maketitle

\section{\label{sec:level1} Introduction}
The decoupling of quantum systems from their environment is of fundamental importance  for realizing quantum states with long coherence times. However, for controlling and measuring the system a controlled interaction with the environment, however weakly, is inevitably required.
Atoms in free space, for example, are decoupled intrinsically well from the radiation field due to their typically small dipole moment. To manipulate their internal electronic state, strong mono- or bi-chromatic laser pulses \cite{Leibfried2003, Haeffner2008} or high-lying Rydberg states with large dipole moments \cite{Raimond2001, Gaetan2009, Evered2023} have to be employed. In contrast, the interaction of superconducting quantum  circuits acting as artificial atoms can be designed at will \cite{Blais2021}. Because of their relatively large coupling strengths the complete and controlled decoupling from external noise sources can then become challenging and in large systems for quantum computing, cross-talk between superconducting qubits is a major hurdle for further scaling \cite{Krinner2020, Acharya2023, Kim2023}. Understanding and circumventing the limits imposed by engineered couplings to control and measure the systems, in particular as the coherence times of superconducting qubits continue to improve through advances in fabrication and qubit design \cite{Place2021, Somoroff2023, Kono2023, Biznarova2023} has, therefore, become imperative. 

For the control of superconducting qubits one typically relies on the weak coupling to control lines and the high signal-to-noise of microwave signal generators. State readout, however, requires a strong coupling between a qubit and a meter system, typically a cavity resonator operated in the dispersive regime, to achieve quantum-limited measurements. 
Here, an effective non-linear dispersive interaction $\propto \hat{a}_q^\dagger \hat{a}_q \hat{a}_r^\dagger \hat{a}_r$ between qubit mode $\hat{a}_q$ and resonator mode $\hat{a}_r$ originates from a linear coupling $\propto \hat{a}_q^\dagger\hat{a}_r + \hat{a}_q\hat{a}_r^\dagger$ when the detuning from the resonator is large compared to the coupling. Since for an efficient and fast measurement the resonator itself needs to be coupled to external degrees of freedom, the linear coupling leads to Purcell-induced decay \cite{Houck2008}. Additionally, residual excitations of the readout mode, e.g., caused by thermal noise or by un-filtered noise from the control electronics, lead to dephasing by randomly shifting the qubit frequency through the dispersive interaction \cite{Schuster2005, Clerk2007, Rigetti2012, Sears2012, Sheldon2017, Yan2018}.

Protection against photon-induced dephasing and Purcell decay can be achieved by controlling the dispersive \cite{Swiadek2023} or directly the linear coupling \cite{Lu2023, Huber2024}.
However, this requires active control of the qubit frequency to change the detuning of the modes, or active control of the magnitude of the linear coupling between qubit and resonator. Both methods introduce additional decoherence channels and, moreover, Purcell-decay cannot be suppressed during readout, when the effective dispersive coupling is maximized.
A further strategy is to shape the density of environmental states of the qubit via additional circuit elements acting as filters between the readout resonator and the transmission line used for probing the system, known as a Purcell filter \cite{Reed2010, Jeffrey2014, Bronn2015, Heinsoo2018, Sunada2022, Sunada2023}. 
Alternatively, one can exploit symmetry properties of quantum states by forming a decoherence free subspace based on the destructive interference of decoherence processes \cite{Lidar1998}.
For example, collective subradiant states \cite{Dicke1954}, realized using trapped ions \cite{Devoe1996, Casabone2015}, cold atoms \cite{Guerin2016} or superconducting qubits \cite{Filipp2011, VanLoo2013, Wang2020, Zanner2022}, or particular states of multi-level atoms \cite{Zhu1996} can exhibit a suppression of spontaneous emission. 
Similarly, superconducting quantum circuits can be used to synthesize artificial multi-level systems, hosting qubit modes protected from intrinsic \cite{Gyenis2021} or environmentally induced decoherence \cite{Gambetta2011}. 
Here, we focus on the latter, implemented using weakly-anharmonic multi-mode circuits, which have also been used for efficient entanglement operations \cite{Roy2017, Roy2018, Hazra2020, Roy2020}, alternative readout schemes \cite{Diniz2013, Dassonneville2020, Dassonneville2022}, two-qubit gates with suppressed cross-talk \cite{Finck2021}, as weakly tunable qubit \cite{Jose2022} or as sensors for charge fluctuations \cite{Wills2022}. 
In particular, a two-mode circuit \cite{Gambetta2011, Srinivasan2011, Hoffman2011, Zhang2017}, has been proposed to protect either against Purcell decay or photon-induced dephasing by controlling the linear or the dispersive coupling to its readout resonator. However, this two-mode qubit design is not able to simultaneously protect against both effects. Moreover, by introducing dynamic tunability of the couplings via flux control, the mode used to encode the qubit becomes sensitive to flux noise. 

To mitigate this effect and allow for simultaneous Purcell and photon-induced dephasing protection of a flux-insensitive qubit mode, we design and characterize a three-mode circuit which includes a fixed-frequency qubit mode that is decoupled from the readout resonator. A second, independently tunable mode is used to mediate an indirect dispersive coupling through a non-linear cross-Kerr coupling between the modes. This indirect dispersive shift can be dynamically controlled by detuning the secondary mode with respect to the readout resonator. Consequently, the qubit mode can simultaneously exhibit negligible linear and dispersive coupling with the readout resonator, protecting it from Purcell decay and photon-induced dephasing while remaining unaffected by flux noise.  

\section{Protected Three-Mode Qubit \label{sec:model}}
\begin{figure}[H]
    \centering
    \includegraphics{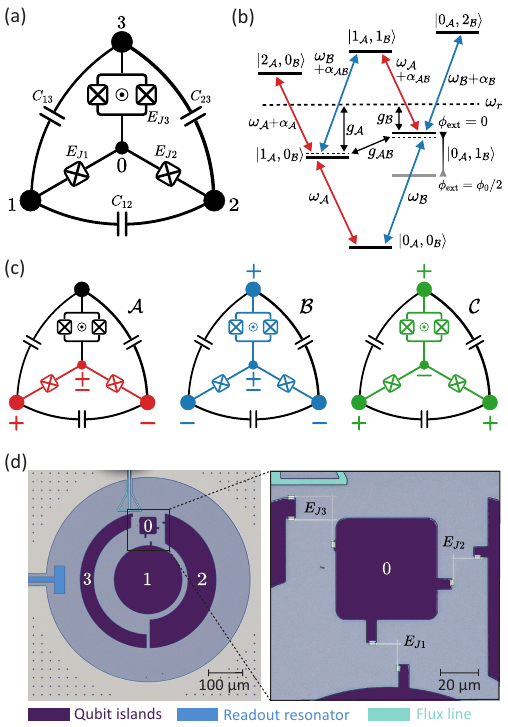}      
 \caption{(a) Circuit diagram with outer islands $1, 2, 3$ and middle island $0$. The most relevant capacitances are $C_{12}, \; C_{13}, \; C_{23}$; capacitances to the center island $C_{0i}$ and to the ground $C_{gi}$ are not explicitly depicted. The Josephson energies between outer and center island are labeled by $E_{Ji}$, with $E_{J3}(\phi_{\rm ext})$ denoting the flux-dependent Josephson energy of the SQUID loop. (b) Energy level diagram restricted to the two-excitation subspace of mode $\mathcal{A}$ and $\mathcal{B}$ relative to the frequency of the readout resonator $\omega_r$. (c) Normal modes under symmetry constraints $C_{13} = C_{23}\neq C_{12}$ and ${E_{J1}=E_{J2}=E_{J3}}$. Fixed frequency mode $\mathcal{A}$ (red) and frequency tunable modes $\mathcal{B}$ (blue) and $\mathcal{C}$ (green). The oscillating charge distributions are indicate by positive ('+') and negative ('-') charge, '$\pm$' indicates charge neutrality. (d) False-color microscope image of the qubit sample including the coupler to the readout resonator (left) and the flux line (top). Islands are labeled as in the circuit diagram in (a), the close up shows the Josephson junctions and SQUID connected to island $0$.}
\label{fig:circuit}
\end{figure}

The underlying idea for the protected three-mode qubit circuit, shown in Fig.~\ref{fig:circuit}(a), is to generate a flux-insensitive qubit mode and a tunable mediator mode. The qubit mode $\mathcal{A}$ is realized by coupling two superconducting islands,  $1$ and $2$,  to a middle island $0$ via Josephson junctions, such that charges oscillate between the outer islands $1$ and $2$, with the middle island remaining charge neutral [Fig.~\ref{fig:circuit}(c)]. Connecting a third island to the middle island via a superconducting interference device (SQUID), introduces two additional flux-tunable modes $\mathcal{B}$ and $\mathcal{C}$. Mode $\mathcal{B}$ exhibits charge oscillations from islands $1$ and $2$ to island $3$ and will be used as a tunable mediator. The remaining mode $\mathcal{C}$, in which charges oscillate between the three outer islands $1$, $2$, $3$ and the center island, can be shifted to high frequencies with a suitable choice of parameters.

Note, that the qubit mode $\mathcal{A}$ remains confined to islands $1$ and $2$ and since no current passes through the SQUID, this mode is flux-insensitive. It remains linearly decoupled from the flux-tunable modes when the islands $1$ and $2$ couple symmetrically to all other circuit elements, i.e., when both the Josephson energies of the junctions to the middle island are equal,  $E_{J1}$=$E_{J2}$, and the capacitances to island $0$ and $3$ are identical, $C_{01}=C_{02}$ and $C_{13}=C_{23}$.
To lift the degeneracy between mode $\mathcal{A}$ and $\mathcal{B}$, an asymmetry $C_{13}=C_{23}\neq C_{12}$ is necessary.

If the effective Josephson energy $E_{J3}(\phi_{\rm ext})$ of the SQUID loop is equal to $E_{J1}, E_{J2}$ the modes $\mathcal{B}$ and $\mathcal{C}$ are decoupled as well, making the modes $\mathcal{A}, \mathcal{B}, \mathcal{C}$ the normal modes of the circuit, as depicted in Fig.~\ref{fig:circuit}(c). Here, $\phi_{\rm ext}$ denotes the magnetic flux threaded through the loop. The non-linearity of the Josephson junctions eventually leads to non-linear cross-Kerr couplings between the modes, which are crucial for mediating the interactions between the qubit mode $\mathcal{A}$ and its environment via the mediator mode $\mathcal{B}$.

For a quantitative analysis of the protected three-mode qubit we write its circuit Hamiltonian
\begin{align}
    \hat{H}_{\rm c} =& \frac{1}{2} \boldsymbol{\hat{q}}^T \bold C^{-1} \boldsymbol{\hat{q}} - \sum_i E_{Ji} \cos\hat{\phi}_i \label{eq:circuit_hamiltonian}
\end{align}
in terms of node fluxes $\hat{\phi}_i$ and charge operators $\hat{q}_i$. The energy scales are determined by the Josephson energies ${E_{Ji}, \; (i=1,2,3)}$ as well as by the charging energies of the islands and their cross-couplings included in the capacitance matrix $\boldsymbol C$. To guarantee protection from charge noise, we choose to operate each relevant mode in the transmon regime \cite{Koch2007,Gambetta2011, Wills2022} by choosing Josephson energies which are large compared to the respective charging energies for each mode (see Appendix \ref{appendix:effective_model}). Approximating the potential to fourth order and expanding the Hamiltonian with respect to modes $\mathcal{A}$, $\mathcal{B}$, $\mathcal{C}$ yields the effective Hamiltonian
\begin{align}
\hat{H}_{\rm q} &= \sum_{m} \omega_m \hat{a}_m^\dagger \hat{a}_m +
\dfrac{\alpha_m}{2} \hat{a}_m^\dagger \hat{a}_m^\dagger \hat{a}_m \hat{a}_m \nonumber \\
& + \sum_{m\neq n} g_{mn} (\hat{a}_{m}^\dagger \hat{a}_{n} + \hat{a}_m \hat{a}_{n}^\dagger) + \alpha_{mn} \hat{a}_m^\dagger \hat{a}_m \hat{a}_{n}^\dagger \hat{a}_{n}, \label{eq:effective_hamiltonian}
\end{align} 
with bosonic operators $\hat{a}_m$, mode frequencies $\omega_m$, mode-mode couplings $g_{mn}$, as well as self- and cross-Kerr terms $\alpha_m, \alpha_{mn}$ with $m, n \in\{\mathcal{A}, \mathcal{B}, \mathcal{C}\}$ (see Appendix \ref{appendix:effective_model}).

For small capacitive couplings of the outer islands to the center island $0$,
${C_{0i} \ll C_{ij}}, i,j = 1,2,3$, mode $\mathcal{C}$ acquires a large charging energy $\propto {C_{0i}}^{-1}$ and consequently $\omega_\mathcal{C} \gg \omega_\mathcal{A}, \omega_\mathcal{B}$. The two relevant low-energy modes $\mathcal{A}$ and $\mathcal{B}$ form a V-shaped energy-level diagram shown in Fig.~\ref{fig:circuit}(b) for the two-excitation subspace with the qubit encoded in the transition $\ket{0_\mathcal{A}, 0_\mathcal{B}} \leftrightarrow \ket{1_\mathcal{A}, 0_\mathcal{B}}$.
Transitions between states of different occupation number of the individual modes are modified by the self-Kerr couplings $\alpha_m$, while the cross-Kerr interaction $\alpha_{\mathcal{A}\mathcal{B}}$ shifts the transitions $\ket{q_\mathcal{A}, 0_\mathcal{B}} \leftrightarrow \ket{q_\mathcal{A}, 1_\mathcal{B}}$ dependent on the qubit-state $q \in \{0,1\}$. 

The exchange-type linear couplings $g_{mn}$ between the modes occur only due to deviations from the considered circuit symmetries.
Introducing an asymmetry in the capacitances $C_{13}$ and $C_{23}$ leads to a linear coupling $g_{\mathcal{A}\mathcal{B}} \propto C_{\Delta,3} = C_{13}-C_{23}$ between mode $\mathcal{A}$ and mode $\mathcal{B}$, which can be used for fast exchange-type operations between these modes with potential applications for the controlled coupling of multiple three-mode qubits. 
In addition, asymmetries in  the Josephson energies, e.g., due to unwanted fabrication or design uncertainties, result in inductive coupling contributions $g_{\mathcal{A}\mathcal{B}}, g_{\mathcal{A}\mathcal{C}} \propto E_{\delta J, 12} = E_{J1} - E_{J2}$.
Inevitably, mode $\mathcal{B}$ and $\mathcal{C}$ interact via a flux-tunable coupling $g_{BC} \propto E_{\Delta J, 123} = 2E_{J3}(\phi_{\rm ext}) - E_{J1} - E_{J2}$. Note that this has no effect on the qubit mode $\mathcal{A}$ if it remains decoupled from the modes $\mathcal{B}$ and $\mathcal{C}$.

\section{Flux-noise in-sensitivity \label{sec:spectrum_t2_flux}}
We realize the protected three-mode qubit in a circular design as shown in Fig.~\ref{fig:circuit}(d). The sample is fabricated on a high-ohmic silicon substrate with niobium ground planes and Al/AlOx/Al Josephson junctions. The qubit capacitances are designed to be in the regime $C_{0i} \ll C_{ij}, \; (i,j = 1, 2, 3)$, such that the modes $\mathcal{A}$ and $\mathcal{B}$ lie at low frequencies around $5~\rm{GHz}$ and mode $\mathcal{C}$ is shifted to a frequency above $20~\rm{GHz}$.
Because of fabrication and design asymmetries in the Josephson energies and capacitances, the uncoupled modes $\mathcal{A}$, $\mathcal{B}$, $\mathcal{C}$ hybridize and we refer to theses dressed modes by $\tilde{\mathcal{A}}$, $\tilde{\mathcal{B}}$, $\tilde{\mathcal{C}}$.
An on-chip flux line controls the magnetic flux $\phi_{\rm ext}$ threading the SQUID loop and is  used to control the mode $\tilde{\mathcal{B}}$ frequency.
For readout and control, the protected three-mode qubit is capacitively coupled to a readout resonator positioned at $\omega_r / 2 \pi = 6.990~\rm{GHz}$ with a coupling $\kappa / 2\pi = 1.32(5)~\rm{MHz}$ to a transmission line. The measured mode frequencies, self- and cross-Kerr terms, dispersive shifts, readout fidelities and single-qubit gate fidelities are summarized in Appendix \ref{appendix:qubit_params}.

\begin{figure}[h]
    \centering
        \includegraphics{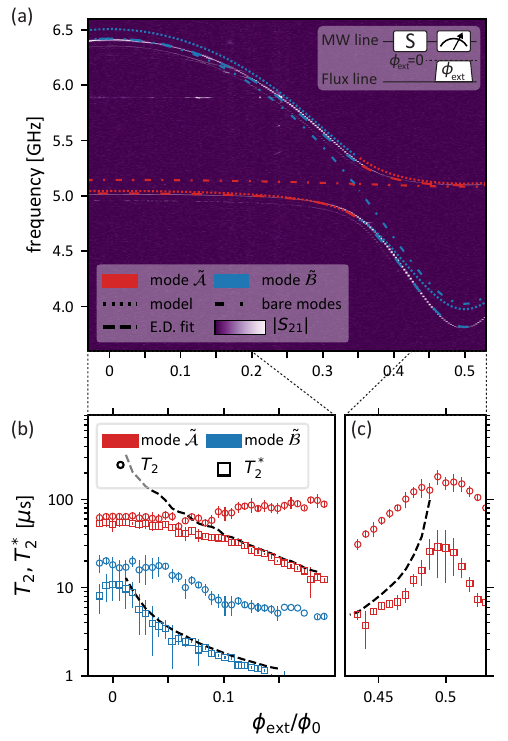}
 \caption{(a) Two tone spectroscopy ($|S_{21}|$) fitted with eigenenergies calculated via exact diagonalization (E.D.) of the circuit model (dashed). Comparison to the effective model with coupled modes (dotted) and bare ($g_{\mathcal{A}\mathcal{B}}=g_{\mathcal{A}\mathcal{C}}=0$) modes (dash-dotted). The inset shows the pulse scheme for control and readout of the qubit. After every pulse sequence (S) on the microwave (MW) line, a flux pulse to $\phi_{\rm ext}=0$ is performed on the flux line prior to readout, to maximize the readout fidelity. (b) and (c) Flux dependence of $T_2$ and $T_2^*$ times of mode $\tilde{\mathcal{A}}$ (red) and $\tilde{\mathcal{B}}$ (blue) at different bias points, compared to the flux noise limit (dashed). The dotted lines between (a), (b) and (c) indicate the part of the spectrum at which the corresponding $T_2, T_2^*$ times are measured.}
 \label{fig:spectrum_t2_vs_flux}
\end{figure}

To characterize the flux dependence of the modes, the spectrum is measured using two-tone spectroscopy \cite{Schuster2005} for different flux-bias values $\phi_{\rm ext}$, as shown in Fig.~\ref{fig:spectrum_t2_vs_flux}(a).
At each point, a flux pulse to $\phi_{\rm ext} = 0$ before the readout maximizes the readout fidelity [Fig.~\ref{fig:spectrum_t2_vs_flux}(a); inset], as further discussed below in Sec. \ref{sec:chi_shifts_purcell_shot_noise}.
Mode $\tilde{\mathcal{A}}$ remains constant with a frequency of $5.017~\rm{GHz}$ at $\phi_{\rm ext} = 0$.
Mode $\tilde{\mathcal{B}}$ is tunable in the range between $3.811$ and $6.408~\rm{GHz}$ and shows an avoided crossing with a coupling strength $g_{\mathcal{A} \mathcal{B}} / 2\pi = 248~\rm{MHz}$ with mode $\tilde{\mathcal{A}}$.
From the measured spectrum, we extract the device parameters by fitting the experimental data to spectra obtained via exact diagonalization of the circuit Hamiltonian in Eq.~(\ref{eq:circuit_hamiltonian}) [Fig.~\ref{fig:spectrum_t2_vs_flux}(a); dashed lines]. For comparison with the numerical solution, the energy spectrum is derived from the effective Hamiltonian in Eq.~(\ref{eq:effective_hamiltonian}) [Fig.~\ref{fig:spectrum_t2_vs_flux}(a); dotted line] based on the obtained circuit parameters. While the numeric solution is in good agreement with the experimental data, the corresponding effective model slightly deviates. This is expected as the approximation is only exact in the transmon regime. However, the effective model can explain the weak flux dependence of mode $\tilde{\mathcal{A}}$ resulting from the mode-mode coupling $g_{\mathcal{A}\mathcal{B}}$, which leads to hybridization with the flux-tunable mode $\mathcal{B}$. The spectrum of the effective Hamiltonian with the couplings $g_{\mathcal{A}\mathcal{B}}$ and $g_{\mathcal{A}\mathcal{C}}$ set to zero, shows the flux-independent nature of the bare mode $\mathcal{A}$ [dotted-dashed lines in  Fig.~\ref{fig:spectrum_t2_vs_flux}~(a)].

To demonstrate the flux-noise insensitivity of the dressed mode $\tilde{\mathcal{A}}$, we compare mode $\tilde{\mathcal{A}}$ and $\tilde{\mathcal{B}}$ by measuring both the $T_2$ and the $T_2^*$ coherence time against the applied external flux $\phi_{\rm ext}$ using standard spin-echo and Ramsey pulse sequences. 
Here, $T_2^{(*)} = (\Gamma_1/2 + \Gamma_\phi)^{-1}$ has contributions from spontaneous decay $\Gamma_1=1/T_1$ and pure dephasing $\Gamma_\phi$.
The flux-noise contribution to $\Gamma_\phi$ is proportional to $|\frac{\partial \omega_m}{\partial \phi_{\rm ext}}|$ \cite{Koch2007} and thus minimal flux noise-sensitivity is expected at flux bias points at which the derivative vanishes, i.e. $\phi_{\rm ext} = 0$ and $\phi_{\rm ext} = \phi_0/2$, and we therefore consider only adjacent regions. 
As shown in  Fig.~\ref{fig:spectrum_t2_vs_flux}(b), mode $\tilde{\mathcal{B}}$ has an inherent strong flux dependence, so its coherence degrades rapidly between $\phi_{\rm ext} = 0$ and $\phi_{\rm ext} = 0.15~\phi_0$ due to flux noise. 
For mode $\tilde{\mathcal{A}}$, we expect a constant $T_2^*$ time when mode $\tilde{\mathcal{B}}$ is sufficiently detuned from mode $\tilde{\mathcal{A}}$. The observed slow decrease in $T_2^*$ between $\phi_{\rm ext} = 0$ and $\phi_{\rm ext} = 0.2 \phi_0$ is predominantly caused by an increased hybridization with the tunable mode $\tilde{\mathcal{B}}$. This effect is enhanced around $\phi_{\rm ext} = \phi_0/2$ as mode $\tilde{\mathcal{B}}$ exhibits stronger flux tuning. Additionally, at $\phi_{\rm ext} = \phi_0/2$, the $T_2^*$ time of mode $\tilde{\mathcal{A}}$ is affected by charge noise. This is because mode $\tilde{\mathcal{B}}$, which hybridizes with mode $\tilde{\mathcal{A}}$, is not completely in the transmon regime.

The $T_2^*$-times are compared to the flux noise limit, which is approximated using the dephasing rate $\Gamma_\phi=\sqrt{A_\phi} \cdot |\frac{\partial \omega_m}{\partial \phi_{\rm ext}}|$ \cite{Koch2007}. Here, a $1/f$-like power spectral density $S_\phi (\omega) = 2\pi A_\phi / |\omega|$ is assumed with a fitted flux noise amplitude $A_\phi = 1.69(8) \; \mu \phi_0$ for a designed SQUID loop size of $210 ~\rm\mu m^2$. The obtained flux noise amplitude $A_\phi$ is consistent with observed values for superconducting qubits \cite{Bylander2011, Hutchings2017, Ganzhorn2020}. 

Both charge and flux-noise typically exhibit a $1/f$-like power spectral density and can therefore be mitigated using a spin-echo pulse sequence. This becomes evident in the measured $T_2$ times [Fig.~\ref{fig:spectrum_t2_vs_flux}(b)], which are much longer than the respective $T_2^*$ times at all flux points where charge- or flux-noise is dominant.

Counter-intuitively, the measured $T_2$-time of mode $\tilde{\mathcal{A}}$ increases when moving away from the flux-insensitive point $\phi_{\rm ext} = 0$, opposed to the corresponding $T_2^*$ times. 
This is caused by the dispersive interaction with the readout resonator, which is controlled by the detuning of mode $\tilde{\mathcal{B}}$ from the resonator. 
At the point $\phi_{\rm ext}=0$, the interaction strength is maximal and thus photon-induced dephasing limits the $T_2$-time. Reducing the dispersive interaction, by moving mode $\tilde{\mathcal{B}}$ away from the resonator, mitigates this effect and consequently leads to an increase in the $T_2$-time.
At $\phi_{\rm ext} = \phi_0/2$, where the longest $T_2$-times are measured, the dispersive interaction is further suppressed, leading to protection from photon-induced dephasing, as discussed in the following.

\section{Protection against Purcell-decay and photon-induced dephasing  \label{sec:chi_shifts_purcell_shot_noise}}
Aside from supporting a flux-insensitive qubit mode, a further advantage of the protected three-mode qubit becomes evident when considering the coupling to other elements. Here we focus on the cross-Kerr mediated interaction between the qubit mode and a readout resonator. A direct linear exchange-type coupling $\propto \hat{a}_q^\dagger \hat{a}_r + h.c.$ would lead to Purcell-enhanced spontaneous emission. In contrast, a cross-Kerr type interaction with an intermediate mode which itself is coupled linearly to the readout resonator can protect the qubit mode from this effect \cite{Gambetta2011}. The mediator mode is able to control a dispersive interaction between the resonator and the qubit mode, which remains linearly decoupled and, therefore, protected against Purcell-decay.
In addition, by reducing the dispersive interaction when detuning the mediator mode from the resonator frequency, photon-induced dephasing can be suppressed at the same time.

In general, the capacitive coupling of the protected three-mode qubit to a readout resonator results in the Hamiltonian
\begin{align}
    \hat{H}_{\rm qr} = \hat{H}_{\rm q} + \omega_r \hat{a}_r^\dagger \hat{a}_r + \sum_m g_{m} (\hat{a}^\dagger_r \hat{a}_m + \hat{a}_r \hat{a}_m^\dagger).\label{eq:qubit_resonator_hamiltonian}
\end{align}
Here, 
\begin{align}
g_{m} = \sqrt{\frac{\hbar}{2Z_r}} \bra{0} \sum_i (\boldsymbol{C}^{-1})_{ir} \hat{q}_i \ket{1_m}
\end{align} 
is the linear coupling between the circuit modes ($m \in \{\mathcal{A}, \mathcal{B}, \mathcal{C}\}$) and the resonator
with an impedance $Z_r$. The coupling $g_m$ depends not only on the capacitive coupling matrix $\boldsymbol{C}$, but also by the matrix element $\bra{0}\hat{q}_i\ket{1_m}$ on the charge distribution of the modes on the islands $i$.   
The characteristic charge distribution of the individual modes indicated in Fig.~\ref{fig:circuit}(c) can be exploited to engineer mode-selective coupling to the readout resonator. 
Coupling capacitively only to circuit island $3$ results in a decoupling of the readout resonator from mode $\mathcal{A}$. In this case the island $3$ remains charge neutral for mode $\mathcal{A}$, and the expectation value $g_\mathcal{A}\propto\bra{0}\hat{q}_3\ket{1_\mathcal{A}}$ is zero, while $g_\mathcal{B}$ remains finite. Note that the same effect can be achieved by symmetrically coupling to islands $1$ and $2$. 

In the dispersive regime $|\Delta_m | = |\omega_m - \omega_r| \gg g_m$ and for single circuit-mode excitations the interaction of the circuit modes with the resonator is described by the dispersive Hamiltonian \cite{Roy2018}  
\begin{align}
\hat{H}_{\rm disp.} = \hat{H}_{\rm q} + \omega_r \hat{a}^\dagger_r\hat{a}_r + \sum_m 2\chi_m\hat{a}^\dagger_r\hat{a}_r \hat{a}^\dagger_m\hat{a}_m,
\end{align}
with the dispersive shifts $\chi_m$ (see Appendix \ref{appendix:coupling_to_resonator}). For the qubit mode $\mathcal{A}$ the dispersive shift is given by
\begin{align}
\chi_\mathcal{A} =& g_\mathcal{A}^2 \left(\frac{1}{\Delta_\mathcal{A}} - \frac{1}{\Delta_\mathcal{A} + \alpha_\mathcal{A}}\right) +  \frac{g_{\mathcal{B}}^2}{2} \left(\frac{1}{\Delta_{\mathcal{B}}} - \frac{1}{\Delta_{\mathcal{B}} + \alpha_{\mathcal{A}\mathcal{B}}}\right). \label{eq:model_chi_shifts}
\end{align}
The first term in Eq.~(\ref{eq:model_chi_shifts}) is the typical \emph{direct} dispersive shift due to the linear coupling $g_\mathcal{A}$. The second term in Eq.~(\ref{eq:model_chi_shifts}) describes the \emph{indirect} dispersive shift mediated by coupling mode $\mathcal{B}$ and its cross-Kerr type interaction $\alpha_{\mathcal{A}\mathcal{B}}$ with the qubit mode. It is caused  by virtual transitions between states with one excitation either in the resonator or the coupling mode, $\ket{1_A,0_B,n=1} \leftrightarrow \ket{1_A,1_B,n=0}$ in second-order perturbation theory
\cite{Zhu2013}. Here, $n$ denotes the resonator photon number. Since the high-frequency $\mathcal{C}$-mode is far detuned its contribution can be neglected.
In the case of a mode-selective coupling with $g_{\mathcal{A}} = 0$ and $g_{\mathcal{B}} \neq 0$, the dispersive interaction $\chi_{\mathcal{A}}$ can be non-zero, mediated exclusively through mode $\mathcal{B}$ and the non-linear cross-Kerr term $\alpha_{\mathcal{A}\mathcal{B}}$ between the modes.

The advantage of our protected three-mode qubit design is that $\chi_\mathcal{A}$ can be controlled via the independently tunable mode $\mathcal{B}$, while the linear coupling $g_\mathcal{A}$ remains suppressed.
This is in contrast to the tunable coupling qubit in Refs.~\cite{Gambetta2011,Srinivasan2011, Hoffman2011,Zhang2017}, in which the dispersive interaction cannot be modified without changing the linear coupling. 
The protected three-mode qubit can therefore be operated in two different regimes of $\chi_{\mathcal{A}}$: the readout regime, where $\chi_{\mathcal{A}}$ is maximal, and the control regime with minimal $\chi_{\mathcal{A}}$ and protection against photon-induced dephasing. In both regimes the qubit mode $\mathcal{A}$ remains Purcell protected and insensitive to flux noise.

We realize the above regimes by capacitively coupling the qubit to the readout resonator mainly via circuit island $3$ to minimize the linear coupling $g_\mathcal{A}$ and to 
suppress the indirect component of $\chi_\mathcal{A}$. While the coupling $g_\mathcal{A}$ can be completely eliminated in the design, we set a weak coupling to allow for qubit control through the readout resonator \cite{Blais2007}. 
The dispersive interaction strength is maximal at the flux-bias value  $\phi_{\rm ext} = 0$ where the detuning of the mode $\tilde{\mathcal{B}}$ from the resonator is minimal. In contrast, for $\phi_{\rm ext} = \phi_0/2$ mode $\tilde{\mathcal{B}}$ is maximally detuned from the resonator and the dispersive interaction is minimized. In the following, we refer to these operation points as readout point $\phi_{\rm r}$ ($\phi_{\rm ext} = 0$) and control point $\phi_{\rm c}$ ($\phi_{\rm ext} = \phi_0/2$), respectively. 
Randomized benchmarking of single-qubit gates consisting of $100~{\rm ns}$ DRAG pulses \cite{Motzoi2009, Werninghaus2021} yields fidelities of $\mathcal{F}_{\rm RB} > 99.9\%$ at both operating points for the qubit mode $\tilde{\mathcal{A}}$. Moreover, the readout fidelity remains consistently high for a $2~\mu{\rm s}$ readout pulse when pulsing to $\phi_\text{r}$ immediately before the readout operation, reducing from $96.3 \%$ at $\phi_\text{r}$ to $94.1\%$ at $\phi_\text{c}$, cf. Appendix \ref{appendix:qubit_params}.

\begin{figure}[h]
    \centering  \includegraphics{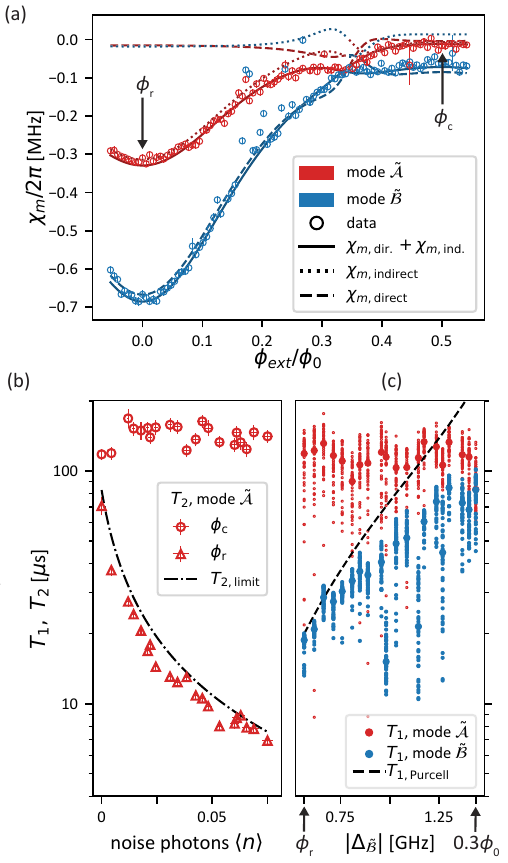}
 \caption{(a) Dispersive interaction $\chi_{\tilde{\mathcal{A}}}, \chi_{\tilde{\mathcal{B}}}$ of modes $\tilde{\mathcal{A}}$ and $\tilde{\mathcal{B}}$ as a function of reduced external flux $\phi_{\rm ext}/\phi_0$. Comparison with sum (solid) of direct (dotted) and indirect (dashed) contributions to the dispersive shifts $\chi_m$ ($m=\tilde{\mathcal{A}},\tilde{\mathcal{B}}$) calculated via Eq.~(\ref{eq:chi_shifts_experiment}). The arrows indicate the readout point $\phi_{\rm r}$ and control point $\phi_{\rm c}$. (b) $T_2$ of mode $\tilde{\mathcal{A}}$ vs. noise photon number at different flux points and comparison to the dephasing limit $T_{2, \rm limit}$ (dash-dotted line). (c) $T_1$ of both modes vs. resonator detuning $\Delta_{\tilde{\mathcal{B}}}$ of mode $\tilde{\mathcal{B}}$ and comparison with the corresponding Purcell limit (dashed line). The large dots represent the median of the individual $T_1$ measurements (small dots). Arrows indicate the flux-bias points that correspond to the realized detunings $\Delta_{\tilde{\mathcal{B}}}$.}
\label{fig:chi_shifts_purcell_shot_noise}
\end{figure}

To confirm the predicted tunability of the couplings, the dispersive shifts of the two modes are measured as a function of applied flux, as shown in Fig.~\ref{fig:chi_shifts_purcell_shot_noise}(a). At each flux point, the dispersive shift is determined by measuring the shift of the resonator frequency with the qubit prepared in its ground state or excited state. The error bars, estimated from the fit of the resonator frequency shift, are comparable to the size of the data points.

Since the transmon approximation is not fully valid over the whole tuning range of mode $\tilde{\mathcal{B}}$, Eq.~(\ref{eq:model_chi_shifts}) is adapted to model the data, see Eq.~(\ref{eq:chi_shifts_experiment}) in Appendix \ref{appendix:coupling_to_resonator}.
The dispersive shifts calculated via equation (\ref{eq:chi_shifts_experiment}) and the experimental data show good agreement. Separating the equation into its direct and indirect part, reveals that the dispersive shift of mode $\tilde{\mathcal{B}}$ is dominated by the direct contribution. In contrast, the small linear coupling $g_{\tilde{\mathcal{A}}}$ and the large detuning of mode $\tilde{\mathcal{A}}$ to the readout resonator suppress the direct dispersive shift contribution. The indirect dispersive shift remains as the dominant contribution.
At the readout point $\phi_{\rm r}$, $\chi_{\tilde{\mathcal{A}}}(\phi_{\rm r}) / 2\pi = -0.32(1)~\rm{MHz}$ is maximal providing the highest readout fidelity (see Appendix \ref{appendix:qubit_params}). At the control point $\phi_{\rm c}$, $\chi_{\tilde{\mathcal{A}}}(\phi_{\rm c}) / 2\pi = -19(4)~\rm{kHz}$ is strongly suppressed, thus protecting mode $\tilde{\mathcal{A}}$ from photon-induced dephasing through residual population of the readout resonator.

To demonstrate the protection against photon-induced dephasing at the control point $\phi_{\rm c}$, we use a broadband noise source, see Appendix \ref{appendix:noise_calibration}, to artificially inject photons into the resonator and track $T_2$ at the flux bias points $\phi_{\rm r}$ and $\phi_{\rm c}$, as shown in Fig.~\ref{fig:chi_shifts_purcell_shot_noise}(b). The average noise photon number is calculated via AC Stark shift measurements at the readout point $\phi_{\rm r}$, $\langle n \rangle = (\omega_{\tilde{\mathcal{A}}} - \omega_{\tilde{\mathcal{A}}, \rm noise}) / (2\chi_{\tilde{\mathcal{A}}})$ with $\omega_{\tilde{\mathcal{A}}, \rm noise}$ denoting the shifted frequency in presence of the additional noise photons. At this point, an increase in noise photons leads to a decrease in $T_2$ according to the photon-induced dephasing rate \cite{Rigetti2012, Clerk2007}
\begin{align}
\Gamma_\phi = \frac{\kappa}{2} \mathrm{Re}\left[ \sqrt{\left(1+ \frac{2 i \chi_{\tilde{\mathcal{A}}}}{\kappa}
\right)^2 + \frac{8 i \chi_{\tilde{\mathcal{A}}} n_{\rm th}}{\kappa}}
- 1 \right], \label{eq:shot_noise_limit}
\end{align}
which limits the coherence to $T_{2,\rm limit} = (\Gamma_1/2 + \Gamma_\phi)^{-1}$.
For the evaluation of $\Gamma_\phi$, we assume an initial thermal population $n_{\rm initial}=0.005$, i.e. the total photon number is given by $n_{th} = n_{\rm initial} + \langle n \rangle$. The resulting estimated $T_{2,\rm limit}$ agrees well with the measured values. 
In contrast, at the control point $\phi_{\rm c}$ the dispersive shift $\chi_{\tilde{\mathcal{A}}} / 2\pi = -19(4)$ kHz is strongly suppressed. Consequently, $T_2$ remains constant over all measured noise photon numbers, thus showcasing the protection against shot noise with an estimated pure dephasing limit of $T_{2, \phi} = 1/\Gamma_\phi = 29(12)~\rm{ms}$.

To demonstrate the Purcell protection of mode $\tilde{\mathcal{A}}$, we measure its $T_1$-time and compare it to mode $\tilde{\mathcal{B}}$ around the readout point $\phi_{\rm r}$ for different values of the detuning $\Delta_{\tilde{\mathcal{B}}}$ of $\tilde{\mathcal{B}}$. Fifty $T_1$ measurements are performed at each value of $\Delta_{\tilde{\mathcal{B}}}$ to capture fluctuations in $T_1$ that are typically observed in superconducting qubits \cite{Klimov2018, Schlor2019, Carroll2022}. 
It is assumed that the $T_1$ fluctuations originate from fluctuating two-level system defects, which we also expected to be the limiting factor for the measured $T_1$-times.
A comparison of both modes along with a Purcell-limit estimate of mode $\tilde{\mathcal{B}}$ given by $T_{1, \rm Purcell} = 1/\kappa (\Delta_{\tilde{\mathcal{B}}}/g_{\tilde{\mathcal{B}}})^2$ is shown in Fig.~\ref{fig:chi_shifts_purcell_shot_noise}(c).
Mode $\tilde{\mathcal{B}}$ is strongly Purcell-limited at the readout point, with increasing detuning the $T_1$-time increases according to the Purcell limit. Deviations from the Purcell limit for larger detunings indicate the presence of another dominating loss channel. In contrast, mode $\tilde{\mathcal{A}}$ maintains a constant $T_1$-time, indicating that Purcell decay into the readout mode is strongly suppressed.

To estimate the Purcell limit on mode $\tilde{\mathcal{A}}$ imposed by the linear couplings $g_{\tilde{\mathcal{A}}}$, we use $g_{\tilde{\mathcal{A}}} / 2\pi = 23.4(2)$~MHz at $\phi_{\rm r}$ and $g_{\tilde{\mathcal{A}}} / 2\pi = 6.5(1)$~MHz at $\phi_{\rm c}$, estimated from circuit parameters extracted from the measured energy spectrum [Fig.~\ref{fig:spectrum_t2_vs_flux}(a)] and dispersive shifts [Fig.~\ref{fig:chi_shifts_purcell_shot_noise}(a)], see Appendix \ref{appendix:coupling_to_resonator}. We find corresponding Purcell limits $T_{1, \rm Purcell} = 1/\kappa (\Delta_{\Tilde{\mathcal{A}}}/g_{\tilde{\mathcal{A}}})^2$ of $855(12)~\mu$s and $10.1(3)~$ms, respectively, well beyond the measured $T_1$-times. While ideally  $g_{\tilde{\mathcal{A}}}$ is constant for varying flux bias values, the coupling between the modes in our device leads to a residual flux dependence.  In fact, the large coupling $g_{\mathcal{B}\mathcal{C}}\propto E_{\Delta J,123}$ between modes $\mathcal{B}$ and $\mathcal{C}$, which is
estimated to vary between $g_{\mathcal{B}\mathcal{C}} / 2\pi = 1.144 \; \rm{GHz}$ at $\phi_{\rm r}$ and $g_{\mathcal{B}\mathcal{C}} / 2\pi = -2.567 \; \rm{GHz}$ at $\phi_{\rm c}$, indirectly influences mode $\tilde{\mathcal{A}}$ because of its hybridization with mode $\mathcal{B}$. 

\section{Conclusion and Outlook}
We have introduced the protected three-mode qubit, a multi-mode circuit which accommodates a qubit mode $\mathcal{A}$ that is insensitive to flux noise and protected from resonator-induced Purcell decay and dephasing caused by residual noise photons. A key element in the realized multi-mode circuit is the formation of a qubit mode that is by design independent of external magnetic fields. A second ingredient is the hierarchy of couplings between the modes. The qubit mode couples ideally only via a cross-Kerr type interaction to a second circuit mode $\mathcal{B}$, acting as a mediator mode to other circuit elements via a linear exchange-type interaction. When coupling to a readout-resonator, the dispersive shift of the qubit mode and, therefore, the strength of the measurement-induced back-action can be adjusted via the independently tunable mediator mode without affecting the $T_1$ time of the qubit.  

Both Purcell protection and the flux noise-insensitivity are fundamentally limited by circuit asymmetries and the resulting hybridization of mode $\mathcal{A}$ and $\mathcal{B}$. In the present device a designed capacitance asymmetry $C_{13} \neq C_{23}$ and fabrication uncertainty-caused Josephson junction asymmetries of $d= (E_{J1}-E_{J2})/(E_{J1}+E_{J2}) = 7\%$ lead to an enhanced mode-mode coupling $g_{\mathcal{A}\mathcal{B}}$. Assuming that a symmetric capacitance $C_{13}=C_{23}$ can be fabricated with high precision and that junction asymmetries can be controlled in future devices via an improved junction fabrication process \cite{Osman2021} or by post-fabrication adjustments of the junction parameters \cite{Hertzberg2021}, the mode-mode coupling can be reduced to $g_{\mathcal{A}\mathcal{B}}/2\pi = 15~\textrm{MHz}$, assuming a spread of $d=1\%$.
Furthermore, the observed sensitivity to charge noise at the control point $\phi_{\rm c}$ can be eliminated in future designs by ensuring the transmon regime over the entire tuning range of mode $\tilde{\mathcal{B}}$.

The readout fidelity can be further improved by increasing the coupling $\kappa$ between the resonator and the transmission line. This requires a larger dispersive interaction, which can be achieved by a smaller detuning of the mediator mode with respect to the readout resonator. Alternatively, the cross-Kerr coupling between qubit and mediator mode can be used directly by resonantly coupling the mediator mode to the readout mode \cite{Diniz2013, Dassonneville2020, Dassonneville2022}. This generates large qubit state-dependent shifts while maintaining the quantum non-demolition character of the measurement.
\begin{figure}[h]
    \centering
        \includegraphics{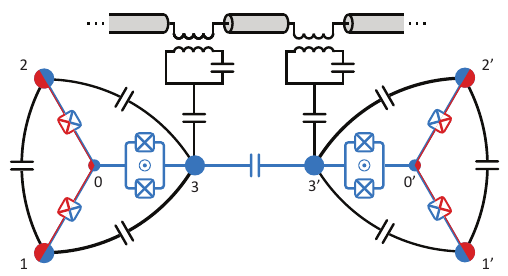}
 \caption{Circuit diagram sketch of two protected three-mode qubits coupled to each other and to their individual readout resonator, which is coupled to a transmission line. The qubit modes (red) are decoupled from each other and their respective readout resonator. The mediator modes (blue) are used to mediate couplings to the corresponding readout resonator and between the two protected-three mode qubits via a mode-selective capacitive coupling.}
 \label{fig:coupled_qubits}
\end{figure}

Similar to the demonstrated tunable indirect dispersive coupling to the readout resonator, the mediator mode $\mathcal{B}$ has the potential to be used for coupling multiple protected three-mode qubits. When coupled to one or more neighbouring protected three-mode qubits, residual interactions between the respective qubit modes can be largely suppressed without the need of further cancellation strategies or passive ZZ cancellation via additional coupling paths \cite{Chen2014, McKay2016, Mundada2019}. 
Mode-selective linear coupling between mediator modes of adjacent three-mode qubits, as illustrated in Fig.~\ref{fig:coupled_qubits}, can be utilized for indirect coupling between protected qubit modes either via linear exchange-type or non-linear cross-Kerr-type interactions with the mediator modes.
Using non-linear couplings only, has the advantage of suppressing spurious cross-talk, originating from stray linear couplings \cite{Billangeon2015, Richer2016}. This has been demonstrated in Ref.~\cite{Finck2021} for two-mode qubits coupled to a common bus resonator.
However, the employed RIP gate \cite{Cross2015} is based on the weak indirect dispersive interaction with the bus resonator, which leads to slow gate times.
Instead, coupled mediator modes $\mathcal{B}$ of adjacent protected three-mode qubits could act as a common element to which the corresponding protected qubit modes $\mathcal{A}$ non-linearly couple via the cross-Kerr interaction $\alpha_{\mathcal{A}\mathcal{B}}$.
Since the cross-Kerr interaction is much larger than the dispersive couplings, this may offer the potential for fast high-fidelity two-qubit gates with suppressed cross-talk.

In summary, with the robustness to coupling-induced incoherent and coherent errors, the protected three-mode qubit is an ideal building block for a quantum processor architecture that retains the performance of a single qubit at large scale. The investigation of architectures with multiple protected qubits and the demonstration of two-qubit gates between the protected modes is a promising future research direction.

\section{Acknowledgements}
We thank Peter Rabl for critically reading the manuscript.
This research was funded by the BMW Group. We acknowledge financial support from the German Federal Ministry of Education and Research via the funding program 'Quantum technologies - from basic research to the market' under contract number 13N15680 (GeQCoS) and under contract number 13N16188 (MUNIQC-SC), by the Deutsche Forschungsgemeinschaft (DFG, German Research Foundation) via project number FI2549/1-1 and via the Germany's Excellence Strategy EXC-2111-390814868 ‘MCQST’ as well as by the European Union with the project grants No. 765267 (QuSCo), 955479 (MOQS) and 847471 (QUSTEC). The research is part of the Munich Quantum Valley, which is supported by the Bavarian state government with funds from the Hightech Agenda Bayern Plus. C.S. has received funding from the European Union's Framework Programme for Research and Innovation Horizon 2020 (2014–2020) under the Marie Sklodowska-Curie Grant Agreement No. 754388 (LMUResearchFellows) and from LMUexcellent.

\appendix
\section{Effective model \label{appendix:effective_model}}
The Lagrangian of the circuit depicted in Fig.~\ref{fig:circuit}(a) is given by
\begin{align}
    \hat{\mathcal{L}} =& \frac{1}{2} \dot{\boldsymbol{\phi}}^T \bold C \dot{\boldsymbol \phi} - \sum_i E_{Ji} \cos\hat{\phi}_i \\
    =& \frac{1}{2} \dot{\boldsymbol \phi}^T \bold C \dot{\boldsymbol \phi} + \frac{1}{2} \boldsymbol \phi^T \bold J \boldsymbol \phi - \sum_i \frac{E_{J,i}}{4!} \hat{\phi}_i^4+ \mathcal{O}(\hat{\phi}_i^6),
\end{align}
with $i \in \{1,2,3\}$ and the capacitance matrix $\bold C$ given by
\begin{align} 
    \begin{pmatrix}
        C_{12} + C_{13} + C_{01} & -C_{12} & -C_{13} \\
        -C_{12} & C_{12} + C_{23} + C_{02} & -C_{23} \\
        -C_{13} & -C_{23} & C_{13} + C_{23} + C_{03}
    \end{pmatrix}.
\end{align}
Here $C_{ij}$ are renormalized capacitances when taking the ground capacitance of each island into account and eliminating the static mode. For simplicity, we set $C_{0i}=C_0$ for all $i \in \{1,2,3\}$ in the following discussion. 
Under the symmetry constraints $C_{13}=C_{23}\neq C_{12}$ and $E_{J1}=E_{J2}=E_{J3}$ the normal modes of the linear Lagrangian, as illustrated in Fig.~\ref{fig:circuit}(c), are given in terms of the node variables by
\begin{align}
\boldsymbol \theta = \bold T \cdot \boldsymbol \phi = \begin{pmatrix}
-1/\sqrt{2} & 1/\sqrt{2} & 0 \\
-1/\sqrt{6} & -1/\sqrt{6} & 2/\sqrt{6} \\
1/\sqrt{3} & 1/\sqrt{3} & 1/\sqrt{3} 
\end{pmatrix} \cdot \boldsymbol \phi,
\end{align}
with the orthonormal transformation matrix $\bold T$. The transformed Lagrangian reads
\begin{align}
    \hat{\mathcal{L}} = \frac{1}{2} \dot{\boldsymbol \theta}^T \boldsymbol{\tilde{C}} \dot{\boldsymbol \theta} + \frac{1}{2} \boldsymbol \theta^T \boldsymbol{\tilde{J}} 
    \boldsymbol \theta - \sum_i E_{J,i} (\sum_j T_{ji} \hat{\theta}_j)^4 + \mathcal{O}(\hat{\phi}_i^6),
\end{align}
with the transformed capacitance 
\begin{align}
\boldsymbol{\tilde{C}} &= \boldsymbol T \cdot \boldsymbol C \cdot \boldsymbol T^{T}  \nonumber \\
&= \begin{pmatrix}
\frac{1}{2} (2C_0 + 4C_{12} + C_{\Sigma,3}) &  \frac{\sqrt{3}}{2} C_{\Delta,3} & 0 \\
\frac{\sqrt{3}}{2} C_{\Delta,3} & C_0+ \frac{3}{2}C_{\Sigma,3} & 0 \\
0 & 0 & C_0
\end{pmatrix} 
\end{align} using
$C_{\Delta,3} = C_{13} - C_{23}$ and $C_{\Sigma,3} = C_{13} + C_{23}$ and the Josephson matrix 
\begin{align}
\boldsymbol{\tilde{J}} &= \boldsymbol T \cdot \boldsymbol J \cdot \boldsymbol T^{T}  \nonumber \\
&= \! \begin{pmatrix}
\frac{1}{2} E_{J,\Sigma} & \frac{1}{2\sqrt{3}} E_{J,\Delta} & -\frac{1}{\sqrt{6}} E_{J,\Delta} \\
\frac{1}{2\sqrt{3}} E_{J,\Delta} & \! \!  \frac{1}{6}(E_{J,\Sigma}+4E_{J3}) & \! \! \! \frac{1}{3\sqrt{2}}(2E_{J3} - E_{J,\Sigma}) \\
-\frac{1}{\sqrt{6}} E_{J,\Delta} & \! \! \frac{1}{3\sqrt{2}}(2E_{J3} -E_{J,\Sigma}) & \! \! \frac{1}{3}(E_{J,\Sigma} + E_{J3})  
\end{pmatrix}\end{align}
with
$E_{J,\Delta} = E_{J1} - E_{J2}$ and $E_{J,\Sigma} = E_{J1} + E_{J2}$. 

A Legendre transformation using $(2e)\bold n = \frac{\partial \mathcal{L}}{\partial \dot{\boldsymbol \theta}} = \tilde{C} \dot{\boldsymbol \theta}$ is applied to obtain the Hamiltonian. Introducing annihilation and creation operators of the harmonic modes
\begin{align}
\hat{\theta}_m &= \left(\dfrac{2E_{C,m}}{E_{J,m}'}\right)^{1/4}(\hat{a}_m^\dagger + \hat{a}_m), \nonumber\\
\hat{n}_m &= \dfrac{i}{2} \left(\dfrac{E_{J,m}'}{2E_{C,m}}\right)^{1/4}(\hat{a}_m^\dagger - \hat{a}_m), \nonumber\\
E_{C,m} &= \dfrac{e^2}{2} (\tilde{C}^{-1})_{mm}, \; E_{J,m}' = \tilde{J}_{mm},
\end{align}
and approximating the potential up to fourth order results in the effective Hamiltonian 
\begin{align}
\hat{H}_{\rm q} &= \sum_m \omega_m \hat{a}_m^\dagger \hat{a}_m + \sum_m
\dfrac{\alpha_m}{2} \hat{a}_m^\dagger \hat{a}_m^\dagger \hat{a}_m \hat{a}_m \nonumber \\
&+ \sum_{m\neq n} g_{mn} (\hat{a}_m^\dagger \hat{a}_n + \hat{a}_m \hat{a}_n^\dagger) + \alpha_{mn} \hat{a}_m^\dagger \hat{a}_m \hat{a}_n^\dagger \hat{a}_n
\label{eq:effective_Hamiltonian_Appendix}
\end{align}
when applying the rotating wave approximation.
The mode frequencies, couplings, self- and cross-Kerr terms are, in this order, given by
\begin{align}
\omega_m &= \omega_{m,0} + \alpha_m + \sum_{m\neq n} \dfrac{\alpha_{mn}}{2}, \; \omega_{m,0} = \sqrt{8E_{C,m} E_{J,m}'},  \nonumber \\
g_{mn} &= e^2 (\tilde{C}^{-1})_{mn} \left(\frac{E_{J,m}'E_{J,n}'}{4E_{C,m}E_{C,n}}\right)^{1/4} \nonumber \\ 
&+ \tilde{J}_{mn} \left(\frac{4E_{C,m} E_{C,n}}{E_{J,m}'E_{J,n}'}\right)^{1/4}, \nonumber \\
\alpha_m &= -\frac{\beta_m}{E_{J,m}'} E_{C,m}, \; \beta_m = E_{J,m} (T_{m1}^4 + T_{m2}^4 + T_{m3}^4), \nonumber \\
\alpha_{mn} &= -2\frac{\gamma_{mn}}{\sqrt{E_{J,m}'E_{J,n}'}}\sqrt{E_{C,m}E_{C,n}}, \nonumber \\
\gamma_{mn} &= \sum_i E_{J,i} T_{mi}^2 T_{ni}^2.
\end{align}
Here the indices $m,n \in \{1,2,3\}$ can be identified with the modes $\mathcal{A}, \mathcal{B}, \mathcal{C}.$
The mode-mode couplings $g_{mn}$ originate from the off-diagonal terms in the transformed capacitance $\boldsymbol{\tilde{C}}$ and Josephson matrix $\boldsymbol{\tilde{J}}$. Therefore, their dependence on the circuit asymmetries can be read off directly from these off-diagonal matrix entries.
In the symmetric case, where $E_{J1}=E_{J2}=E_{J3}$, the self- and cross-Kerr terms simplify to
\begin{align}
    \alpha_\mathcal{A} &= -\frac{E_{C,\mathcal{A}}}{2}, \; \alpha_\mathcal{B} = -\frac{E_{C,\mathcal{B}}}{2}, \; \alpha_\mathcal{C} = -\frac{E_{C,\mathcal{C}}}{3}, \nonumber \\
    \alpha_{\mathcal{A}, \mathcal{B}} &= -\frac{1}{3} \sqrt{E_{C,\mathcal{A}} E_{C,\mathcal{B}}}, \; \alpha_{\mathcal{A}, \mathcal{C}} = -\frac{2}{3} \sqrt{E_{C,\mathcal{A}} E_{C,\mathcal{C}}}, \nonumber \\
    \alpha_{\mathcal{B}, \mathcal{C}} &= -\frac{2}{3} \sqrt{E_{C,\mathcal{B}} E_{C,\mathcal{C}}},
\end{align}
showing that the self-Kerr terms of mode $\mathcal{A}$ and $\mathcal{B}$ are diminished by a factor of 2, compared to the conventional transmon self-Kerr term.

\section{Coupling to the resonator \label{appendix:coupling_to_resonator}}
When capacitively coupling the multi-mode circuit described by the circuit Hamiltonian $\hat{H}_c$ in Eq.~(\ref{eq:circuit_hamiltonian}) to a harmonic readout resonator, we obtain the Hamiltonian
\begin{align}
\hat{H}_{\rm qr} = \hat{H}_{\rm c} + \omega_r \hat{a}_r^\dagger \hat{a} + \sum_{k=1}^3 (\boldsymbol{C}^{-1})_{kr} \hat{q}_k \hat{q}_r\label{eq:qubit_resontor_H_circuit}
\end{align}
with resonator frequency $\omega_r$, capacitance matrix $\boldsymbol{C}$, and resonator charge operator $\hat{q}_r = i \sqrt{\hbar / 2Z_r}(\hat{a}_r^\dagger - \hat{a}_r)$, where $Z_r$ is the resonator impedance. Expressing the Hamiltonian $\hat{H}_{\rm qr}$ in terms of the eigenstates and eigenvalues $\{\ket{i}, \omega_i\}$ of  $\hat{H}_{\rm c}$ results in:
\begin{align}
\hat{H}_{\rm qr} =& \sum_i \omega_i \ket{i}\bra{i} + \omega_r \hat{a}_r^\dagger \hat{a}_r + \sum_{ij} g_{i,j} \ket{i}\bra{j}(\hat{a}_r^\dagger - \hat{a}_r), \nonumber \\
g_{i,j} =& i\sqrt{\hbar/2Z_r}\sum_{k=1}^3 (\boldsymbol{C}^{-1})_{kr}\bra{i}\hat{q}_k\ket{j}. \label{eq:qubit_resonator_numeric}
\end{align}
This has the general form of a multi-level system coupled to a harmonic mode. In the dispersive regime $|\omega_i - \omega_j - \omega_r| \gg |g_{i,j}|$ the Hamiltonian reduces to \cite{Zhu2013}
\begin{align}
\hat{H}_{\rm disp} &= \omega_r \hat{a_r}^\dagger \hat{a}_r + \sum_j (\omega_j +\Lambda_j) \ket{j}\bra{j}  + \sum_j \chi_j \hat{a}^\dagger \hat{a} \ket{j}\bra{j}, \nonumber \\
\Lambda_j &= \sum_i \chi_{ij}, \; \chi_j = \sum_i (\chi_{ij} - \chi_{ji}), \; \chi_{ij} = \dfrac{|g_{i,j}|^2}{\omega_j-\omega_i-\omega_r}. \label{eq:H_disp_general}
\end{align}
From the eigenstates and eigenvalues $\{\ket{i}, \omega_i\}$, as obtained from exact diagonalization of the circuit Hamiltonian $\hat{H}_{\rm c}$ in the charge basis, the couplings $g_{i,j}$ and corresponding dispersive shifts $\chi_i$ can be evaluated numerically.
When instead approximating the circuit Hamiltonian $\hat{H}_{\rm c}$ using the effective Hamiltonian $\hat{H}_q$ in Eq.~(\ref{eq:effective_Hamiltonian_Appendix}), and expressing $\hat{q}_k$ in terms of the mode operators $\hat{a}_m$ in Eq.~(\ref{eq:qubit_resontor_H_circuit}), the resonator-qubit Hamiltonian $\hat{H}_{qr}$ can be expressed in the form 
\begin{align}
    \hat{H}_{\rm qr} = \hat{H}_{\rm q} + \omega_r \hat{a}_r^\dagger \hat{a}_r + \sum_m g_{m} (\hat{a}^\dagger_r \hat{a}_m + \hat{a}_r \hat{a}_m^\dagger),
\end{align}
with $g_m = g_{0,m}$ and $m = \mathcal{A}, \mathcal{B}, \mathcal{C}$, equivalent to the Hamiltonian in Eq.~(\ref{eq:qubit_resonator_hamiltonian}).
In the dispersive regime $|\omega_m - \omega_r| \gg g_m$ the Hamiltonian reduces to \cite{Roy2018}
\begin{align}
\hat{H}_{\rm disp.} = \hat{H}_{\rm q} + \omega_r \hat{a}^\dagger_r\hat{a}_r + \sum_m 2\chi_m\hat{a}^\dagger_r\hat{a}_r \hat{a}^\dagger_m\hat{a}_m,
\end{align}
in the case of a single circuit mode excitation, with the dispersive shifts $\chi_m$ given by
\begin{align}
\chi_m =& g_m^2 \left(\frac{1}{\Delta_m} - \frac{1}{\Delta_m + \alpha_m}\right) + \nonumber \\
&\sum_{n\neq m} \frac{g_{n}^2}{2} \left(\frac{1}{\Delta_{n}} - \frac{1}{\Delta_{n} + \alpha_{mn}}\right),
\end{align}
with $n, m \in \{\mathcal{A}, \mathcal{B}, \mathcal{C}\}$. Note, the above dispersive shifts can be derived from the general expression in Eq.~(\ref{eq:H_disp_general}), by applying
selection rules to the couplings $g_{i,j}$, which are present in the transmon regime \cite{Koch2007} with additional contributions in case of multi-mode circuits \cite{Gambetta2011, Roy2018}. When not operating in the transmon regime, the relations $g_{m,2m} = \sqrt{2}g_{0,m}$ and $g_{n,mn} = g_{0,m}$ are no longer valid and the dispersive shifts are more precisely given by 
\begin{align}
\chi_m =& g_m^2 \frac{1}{\Delta_m} - \frac{g_{m, 2m}^2}{2}\frac{1}{\Delta_m + \alpha_m} + \nonumber \\
&\sum_{n\neq m} \frac{g_{n}^2}{2} \frac{1}{\Delta_{n}} - \frac{g_{n, mn}^2}{2}\frac{1}{\Delta_{n} + \alpha_{mn}}.
\label{eq:chi_shifts_experiment}
\end{align}
This expression we use for modeling the experimentally measured dispersive shifts shown in Fig.~\ref{fig:chi_shifts_purcell_shot_noise}(a). The self- and cross-Kerr terms are calculated via exact diagonalization of the circuit Hamiltonian $\hat{H}_{\rm c}$  Eq.~(\ref{eq:circuit_hamiltonian}), for which we include the full capacitive coupling matrix of the qubit and the resonator. The obtained eigenstates and eigenvalues, are then used to numerically evaluate the couplings using Eq.~(\ref{eq:qubit_resonator_numeric}).
Here, the capacitance matrix and the Josephson energies are estimated by fitting the circuit model to the experimentally measured spectrum [see Fig.~\ref{fig:spectrum_t2_vs_flux}(a)] and the dispersive shifts at $\phi_{\rm ext} = 0$ using a least-mean-squared cost-function and an numeric optimizer provided by SciPy \cite{SciPy2020}.
To estimate the uncertainties of the fitted model parameters, we sample 100 artificial data sets within the initial fit residuals and then repeat the optimization for each data set. This yields a distribution of each model parameter with a corresponding standard deviation and is know as bootstrapping \cite{Bradley1994}. 

\section{Device parameters \label{appendix:qubit_params}}
The device parameters at the readout and control point are summarized in Table~\ref{tab:qubit_params}. 
\begin{table}[H]
\caption{\label{tab:qubit_params}%
Parameters of mode $\tilde{\mathcal{A}}$ and $\tilde{\mathcal{B}}$ at the readout point $\phi_{\rm r}$ ($\phi_{\rm ext} = 0$) and control point $\phi_{\rm c}$ ($\phi_{\rm ext} = \phi_0/2$). The mode frequency $\omega_m$ is measured via a Ramsey sequence, the self-Kerr terms $\alpha_m$, cross-Kerr couplings $\alpha_{\tilde{\mathcal{A}}\tilde{\mathcal{B}}}$ and dispersive shift $\chi_m$ are measured in spectroscopy. The single qubit gate fidelity $\mathcal{F}_{\rm RB}$ is obtained via randomized benchmarking (RB).  The single-shot measurement fidelity is given by $\mathcal{F}_{\rm meas.}$.
}
\begin{ruledtabular}
\begin{tabular}{ccccc}
& $\tilde{\mathcal{A}}(\phi_{\rm r})$ & $\tilde{\mathcal{B}}(\phi_{\rm r})$ & $\tilde{\mathcal{A}}(\phi_{\rm c})$ & $\tilde{\mathcal{B}}(\phi_{\rm c})$ \\
\colrule
$\omega_m/2\pi$ (GHz) & 5.017 & 6.408 & 5.101 & 3.811 \\
$\alpha_m/2\pi$ (MHz) & -117 & -111 & -98 & -403 \\
$\alpha_{\tilde{\mathcal{A}}\tilde{\mathcal{B}}} / 2\pi$ (MHz) & -72 & -72 & -119 & -119 \\
$\chi_m / 2\pi$ (MHz) & -0.32(1) & -0.70(1) & -0.019(4) & -0.05(1) \\
$\mathcal{F}_{\rm RB}$ (\%) & 99.933(2) & 99.844(7) & 99.948(3) &  \\
$\mathcal{F}_{\rm meas.}$ (\%) & 96.3 & 94.2 & 94.1 & 
\end{tabular}
\end{ruledtabular}
\end{table}

\begin{figure}[H]
    \centering
        \includegraphics{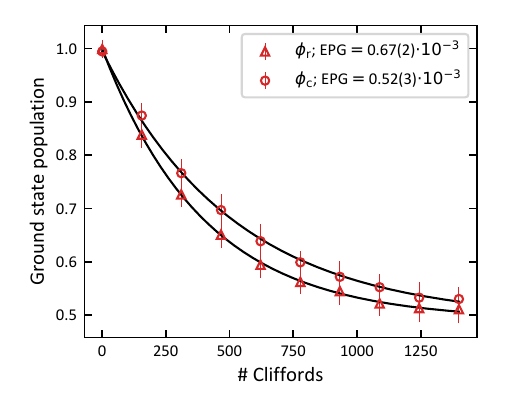}
 \caption{Comparison of single-qubit gate randomized benchmarking (RB) of mode $\tilde{\mathcal{A}}$ at the readout point $\phi_{\rm r}$ and control point $\phi_{\rm c}$ and corresponding error per gate (EPG).}
 \label{fig:device_params}
\end{figure}

Note, that the self-Kerr terms are reduced by a factor of two compared to a conventional transmon qubit \cite{Koch2007}, as expected from the model described in Appendix \ref{appendix:effective_model}. The use of optimal control \cite{Motzoi2009, Werninghaus2021} still allows high-fidelity single-qubit gates as determined by a randomized benchmarking sequence with $100~{\rm ns}$ DRAG pulses resulting in fidelities of $\mathcal{F}_{\rm RB} > 99.9$\% at both operating points for the qubit mode $\tilde{\mathcal{A}}$ [see Fig.~\ref{fig:device_params}].  
The readout fidelities shown in Tab. \ref{tab:qubit_params} are determined for a $2~\mu{\rm s}$ readout pulse.
The single-qubit gate fidelity and the readout fidelity of mode $\tilde{\mathcal{B}}$ at the control point has not been determined, since it is charge noise limited at this operating point.

\begin{figure}[H]
    \centering
        \includegraphics{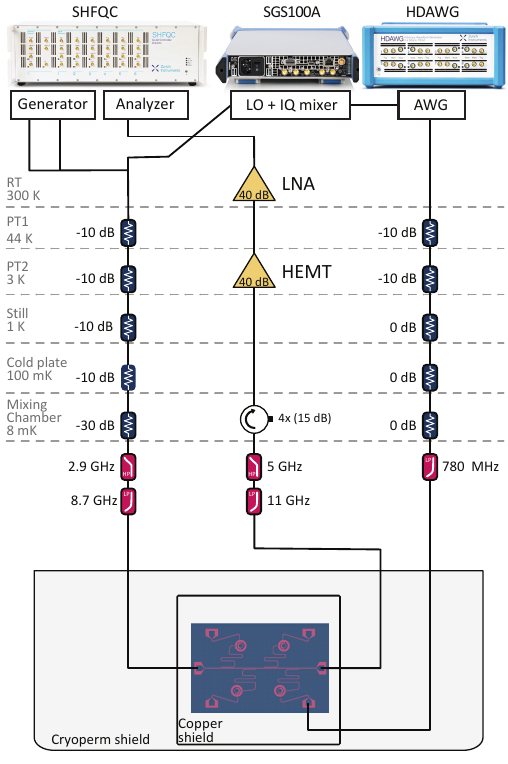}
 \caption{Wiring diagram of the experimental setup.}
 \label{fig:exp_setup}
\end{figure}
\section{Experimental setup}
The wiring diagram of the experimental setup is shown in Fig.~\ref{fig:exp_setup}. Measurements are performed in a BlueFors XLD1000 dilution refrigerator with a base temperature of 8 mK. 
A common input microwave line is used for readout and qubit control. Microwave pulses are generated using a Zurich Instruments SHFQC qubit controller and a Zurich Instruments HDAWG arbitrary waveform generator in combination with a Rhode \& Schwarz SGS100A local oscillator (LO) with integrated mixer, which is able to provide more input power than the SHFQC and is also used as the noise source.
The on-chip flux line is controlled via a Zurich Instruments HDAWG arbitrary waveform generator. Microwave attenuators and filters provide isolation from thermal noise and noise from the room temperature signal sources. The readout signal is amplified using a LNF-LNC4\_8F high electron mobility transistor (HEMT) and BZ-04000800-081045-152020 low noise amplifier (LNA). Single shot readout is performed by the Zurich Instruments SHFQC waveform generator and analyzer.  

\section{Noise photon calibration \label{appendix:noise_calibration}}
\begin{figure}[H]
    \centering
        \includegraphics{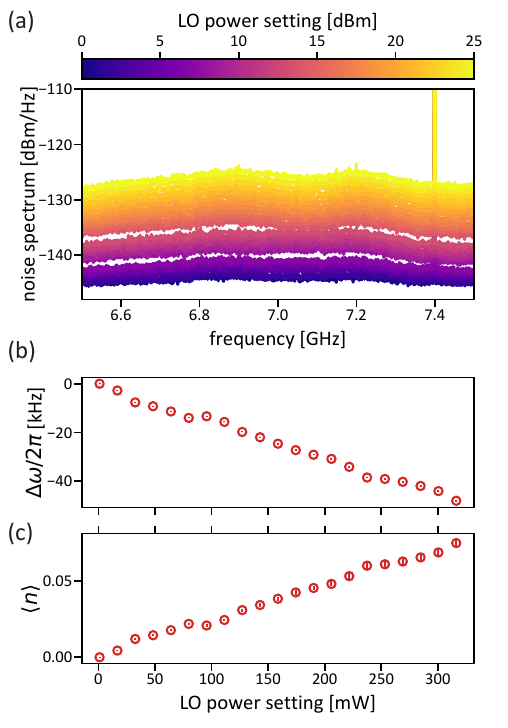}
 \caption{(a) Noise power spectrum of the output signal of the Rhode \& Schwarz SGS100A LO at room temperature for different power settings. (b) Frequency shift $\Delta \omega$ of mode $\tilde{\mathcal{A}}$ at the readout point $\phi_{\rm{r}}$ in dependence of the LO power setting. (c) Average noise photon number $\langle n \rangle$ as a function of the LO power setting.}
 \label{fig:noise_generation}
\end{figure}

We use the power-dependent background noise of the Rhode \& Schwarz SGS100A LO to artificially populate the readout resonator with noise photons.
The noise power spectral density of the output signal of the LO is measured at room temperature as a function of its power setting, as shown in Fig.~\ref{fig:noise_generation}(a), using a spectrum analyzer (Rhode \& Schwarz FSW26). It exhibits a broadband background noise spectrum around the readout resonator frequency $f_r=6.990~\rm{GHz}$ with a peak at the set LO frequency of $7.4~\rm{GHz}$. The background noise power spectral density increases with the LO power setting and leads to a population of the readout resonator. Note, that measurements using different LO frequency settings, which are not presented here, show that the resonator population is indeed caused by the background noise spectrum and not by the LO resonance. To calibrate the average number of noise photons, we measure the shift of the qubit frequency $\Delta \omega = \omega_{\tilde{\mathcal{A}}} - \omega_{\tilde{\mathcal{A}}, \rm noise}$ of mode $\tilde{\mathcal{A}}$ at the readout point $\phi_{\rm r}$ ($\phi_{\rm ext} = 0$) for different LO power settings [Fig.~\ref{fig:noise_generation}(b)], with $\omega_{\tilde{\mathcal{A}}, \rm noise}$ denoting the shifted frequency in the presence of the noise. The error bars are given by the fit of the qubit frequency, which is measured in spectroscopy, and are small compared to the size of the data points. The average noise photon number is calculated via $\langle n \rangle = \Delta \omega / (2\chi_{\tilde{\mathcal{A}}})$ from the measured dispersive shift $\chi_{\tilde{\mathcal{A}}} / 2\pi =-0.32(1)~\rm{MHz}$ at the readout point $\phi_{\rm{r}}$, as shown in Fig.~\ref{fig:noise_generation}(c).  
By measuring the coherence time $T_2$ for different LO power settings, we obtain the $T_2$-time dependent on the noise photon number, as shown in Fig.~\ref{fig:chi_shifts_purcell_shot_noise}(b) in the main text.

\hypersetup{urlcolor=blue}
%

\end{document}